\documentclass[prc,floatfix,groupedaddress,nofootinbib,showpacs,preprintnumbers,
amsmath,amssymb,amsfonts,superscriptaddress,widetable] {revtex4-1}
\usepackage{graphicx}
\usepackage{dcolumn}
\usepackage{mathrsfs}
\usepackage{bm}
\usepackage{float}
\usepackage{soul}
\usepackage[usenames]{color}

\newcommand{\rhoz}{\rho_{\raisebox{-0.75pt}{\tiny 0}}}
\newcommand{\epsz}{\varepsilon_{\raisebox{-0.75pt}{\tiny 0}}}

\newcommand{\gammai}[1]{\gamma^{\raisebox{-1.0pt}{\scriptsize{#1}}}}

\begin{document}

\title{Emergence of low-energy monopole strength \\ in the neutron-rich calcium isotopes}
\author{J. Piekarewicz}
\email{jpiekarewicz@fsu.edu}
\affiliation{Department of Physics, Florida State University, Tallahassee, FL 32306}
\date{\today}
\begin{abstract}
\begin{description}
\item[Background] The isoscalar monopole response of neutron-rich
nuclei is sensitive to both the incompressibility coefficient of symmetric
nuclear matter and the density dependence of the symmetry energy. 
For exotic nuclei with a large neutron excess, a low-energy component 
emerges that is driven by transitions into the continuum.
\item[Purpose] While understanding the scaling of the giant monopole
resonance with mass number is central to this work, the main goal of
this paper is to explore the emergence, evolution, and origin of low energy 
monopole strength along the even-even calcium isotopes: from ${}^{40}$Ca 
to ${}^{60}$Ca.
\item[Methods] The distribution of isoscalar monopole strength is
computed in a relativistic random phase approximation (RPA) using 
three effective interactions that have been calibrated to the properties 
of finite nuclei and neutron stars. A non-spectral approach is adopted 
that allows for an exact treatment of the continuum without any 
reliance on discretization. This is particularly critical in the case of
weakly-bound nuclei with single-particle orbits near the continuum.
The discretization of the continuum is neither required nor admitted. 
\item[Results] For the stable calcium isotopes, no evidence of low-energy 
monopole strength is observed, even as the 1f$^{\,7/2}$ neutron orbital is 
being filled and the neutron-skin thickness progressively grows. Further, in
contrast to experimental findings, a mild softening of the monopole response
with increasing mass number is predicted. Beyond ${}^{48}$Ca, a significant 
amount 
of low-energy monopole strength emerges as soon as the weak-binding 
neutron orbitals (2p and 1f$^{\,5/2}$) become populated. The emergence 
and evolution of low-energy strength is identified with transitions from 
these weakly-bound states into the continuum---which is treated exactly 
in the RPA approach. Moreover, given that models with a soft symmetry
energy tend to reach the neutron-drip line earlier than their stiffer 
counterparts, we identify an inverse correlation between the 
neutron-skin thickness and the inverse energy weighted sum. 
\item[Conclusions] Despite experimental claims to the contrary, we found
a mild softening of the giant monopole resonance in going from ${}^{40}$Ca 
to ${}^{48}$Ca. Measurements for other stable calcium isotopes may be critical 
in elucidating the nature of the discrepancy. Moreover, given the early success 
in measuring the distribution of isoscalar monopole strength in the unstable 
${}^{68}$Ni nucleus, we advocate new measurements along the unstable 
neutron-rich calcium isotopes to explore the critical role of the continuum 
in the development of a soft monopole mode.

\end{description}
\end{abstract}

\maketitle

\section{Introduction}
\label{intro}

What are the relatively few combinations of neutrons and protons that 
form a bound atomic nucleus is a question of critical importance to both 
nuclear structure and astrophysics\,\cite{LongRangePlan,NuPECC2017}. 
Probing the limits of nuclear existence lies at the heart of the commissioning 
of state of the art rare isotope facilities throughout the world. However, mapping 
the precise boundaries of the nuclear landscape is enormously challenging. 
Given that the Coulomb interaction severely limits the number of neutrons that 
may be removed from a given isotope, the proton drip lines remains
relatively close to the valley of stability. Indeed, the proton drip line has 
been determined experimentally up to protactinium, an isotope with atomic 
number $Z\!=\!91$. In stark contrast, the neutron drip line has been mapped 
only up to oxygen\,\cite{Thoennessen:2004}. Thus, the neutron-rich landscape 
constitutes a fertile ground for research in nuclear structure and---due to 
its relevance to the r-process and to the composition of the neutron-star 
crust---also in astrophysics\,\cite{Utama:2017wqe}.

Exceptional experimental advances have led to a paradigm shift in 
fundamental core concepts that have endured the test of time, such 
as the traditional nuclear magic numbers. Understanding the impact 
of such a paradigm shift in the development of novel modes of excitation
in exotic nuclei has become 
a fruitful area of research\,\cite{Paar:2007bk}. Given the richness of 
these excitations, they offer a unique window into the 
nuclear dynamics that is often closed to other means\,\cite{Harakeh:2001}. 
In particular, the isovector dipole resonance has been shown to be highly 
sensitive to the equation of state of neutron-rich matter, specifically to the 
density dependence of the symmetry energy\,\cite{Reinhard:2010wz,
Piekarewicz:2010fa,Piekarewicz:2012pp,Roca-Maza:2013mla}. Indeed, 
measurements of the electric dipole polarizability in a variety of stable 
neutron-rich nuclei\,\cite{Tamii:2011pv,Poltoratska:2012nf,
Hashimoto:2015ema,Birkhan:2016qkr} and in the unstable 
${}^{68}$Ni isotope\,\cite{Wieland:2009,Rossi:2013xha}, provide an 
attractive alternative to parity-violating experiments to determine the 
neutron-skin thickness of ${}^{208}$Pb\,\cite{Abrahamyan:2012gp,
Horowitz:2012tj} as well as the density dependence of the 
symmetry energy.

Of specific interest in this paper is the emergence and evolution of isosocalar 
monopole strength along the calcium isotopes. There are several reasons for 
undertaking such a study. First, the compressibility of neutron-rich matter is 
sensitive to both the incompressibility coefficient of symmetric nuclear matter 
and the density dependence of the symmetry energy\,\cite{Piekarewicz:2008nh}. 
Particularly interesting is to examine this sensitivity along an isotopic chain having 
several stable isotopes, such as calcium and tin. Note that although a decade 
has already passed since the ``fluffiness" of tin was first 
identified\,\cite{Garg:2006vc,Li:2007bp,Li:2010kfa}, a theoretical explanation 
continues to elude us\,\cite{Piekarewicz:2007us,Sagawa:2007sp,Avdeenkov:2008bi,
Li:2008hx,Piekarewicz:2009gb,Khan:2009xq,Khan:2009ih,Khan:2010mv,
Cao:2012dt,Vesely:2012dw,Piekarewicz:2013bea,Colo:2013yta,
Chen:2013tca,Chen:2013jsa}. That is, theoretical models that account for 
Giant Monopole Resonance (GMR) energies in ${}^{90}$Zr, ${}^{144}$Sm, and 
${}^{208}$Pb, overestimate the GMR energies in all stable tin---and 
cadmium ($Z\!=\!48$)\,\cite{Patel:2012zd}---isotopes. The imminent report
of the GMR energy of the unstable ${}^{132}$Sn nucleus may prove vital in 
elucidating the softness of tin\,\cite{Garg:2017PC}. In turn, an exploration of 
the evolution of the GMR energies along the isotopic chain in calcium above 
and beyond the already measured ${}^{40}$Ca and ${}^{48}$Ca 
isotopes\,\cite{Youngblood:2001mq,Lui:2011zz,Anders:2013rea} may be 
highly illuminating. This is particularly relevant given that existing experimental
data seem to suggest an increase in the GMR energies with increasing 
mass, a fact that is difficult to reconcile with theoretical 
expectations\,\cite{Anders:2013rea}. Second, theoretical and computational 
advances have evolved to such an extent that ab initio calculations in the calcium 
region are now possible. Ab initio calculations now exist that can predict 
the electric dipole polarizability as well as the charge and weak form factors of 
both ${}^{40}$Ca and ${}^{48}$Ca\,\cite{Hagen:2015yea}. Indeed, a primary
motivation for the Calcium Radius EXperiment (``CREX") at Jefferson Lab
is to bridge ab initio approaches with nuclear density functional theory\,\cite{CREX:2013}.
Finally, extending GMR calculations beyond ${}^{48}$Ca highlights the role 
of the continuum in the emergence of low-energy monopole strength. We expect 
that transitions out of the weakly bound $pf$ neutron orbitals dominate the 
development of a soft monopole mode.

Although the compressibility of neutron-rich matter is primarily sensitive to the 
incompressibility coefficient of symmetric nuclear matter, a sensitivity to the 
symmetry energy develops in heavy nuclei with a significant neutron excess. 
Unfortunately, this sensitivity is often hindered by the relatively small neutron 
excess of the stable nuclei investigated up to date; note that the contribution 
from the symmetry energy scales as the \emph{square} of the neutron-proton 
asymmetry 
$\alpha\!\equiv\!(N\!-\!Z)/A$\,\cite{Piekarewicz:2008nh}. Hence, measuring the 
distribution of isoscalar monopole strength in unstable nuclei with a large neutron 
excess is highly appealing. Despite the relatively modest value of  
$\alpha_{68}\!=\!0.18$ (and $\alpha_{68}^{2}\!=\!0.03$) the measurement of
the isoscalar monopole response of ${}^{68}$Ni by Vandebrouck and 
collaborators represents an important first step in the right 
direction\,\cite{Vandebrouck:2014}. Although this pioneering experiment 
established the great potential of inelastic alpha scattering in inverse 
kinematics to probe the distribution of isoscalar monopole strength in 
unstable neutron-rich nuclei, a controversy ensued on the interpretation 
of the observed low-energy structure. Based on RPA predictions that used 
a discretized continuum, it was suggested that the observed low-energy 
strength consisted of individual peaks that were well separated from the 
main giant resonance\,\cite{Capelli:2009zz,Khan:2011ej}. Shortly after, 
Hamamoto and Sagawa concluded based on Skyrme-RPA calculations 
that did not rely on a discretization of the continuum, that the 
development of isoscalar monopole peaks in the low-energy region 
was very unlikely\,\cite{Hamamoto:2014ala}. Subsequently, relativistic 
RPA calculations confirmed that while significant amount of low-energy 
strength in ${}^{68}$Ni is indeed generated, such low-energy structure 
lacks any distinct features\,\cite{Piekarewicz:2014bva}. These findings 
suggest that a proper treatment of the continuum is critical. As we shall 
see below, the continuum plays a fundamental role in the development 
of a soft monopole mode in all unstable neutron-rich calcium isotopes.

The paper has been organized as follows. In Sec.\,\ref{Formalism}
a brief summary of the relativistic RPA formalism is presented
highlighting the role of the continuum. In Sec.\,\ref{Results} 
we display both uncorrelated and RPA predictions for the distribution 
of isoscalar monopole strength along the isotopic chain in calcium
for all even-even isotopes: from ${}^{40}$Ca up to ${}^{60}$Ca. In
particular, we pay special attention to the role of the continuum in 
shaping the distribution of low-energy strength. Finally, we provide 
a summary of the most important findings and suggestions for the 
future in Sec.\,\ref{Conclusions}.

\section{Formalism}
\label{Formalism}

The main goal of this section is to provide the technical material 
necessary to follow the discussion presented in Sec.\,\ref{Results}.
The relativistic RPA formalism as implemented here has been 
reviewed extensively in earlier publications; see for example
Refs.\,\cite{Piekarewicz:2013bea,Piekarewicz:2014bva} and 
references contained therein. In particular, the critical role that
the continuum plays on the emergence of low-energy isoscalar 
monopole strength has been extensively addressed in
Ref.\,\cite{Piekarewicz:2014bva}.

\subsection{Isoscalar Monopole Response}
\label{IsMR}

In the relativistic mean-field (RMF) approach pioneered by Serot and
Walecka\,\cite{Walecka:1974qa,Serot:1984ey} the basic constituents 
of the effective theory are protons and neutrons interacting via the
exchange of the photon and various mesons of distinct Lorentz and
isospin character. Besides the conventional Yukawa couplings of the
mesons to the relevant nuclear currents,  the model is supplemented 
by a host of nonlinear meson coupling that have
been found essential to improve the standing of the 
model\,\cite{Boguta:1977xi,Mueller:1996pm,Horowitz:2000xj}.  In the 
widely used mean-field approximation, the nucleons satisfy a Dirac 
equation in the presence of strong scalar and vector potentials that
are the hallmark of the relativistic approach. In turn, the photon and
the mesons satisfy classical Klein-Gordon equations with the 
relevant nuclear densities acting as source term. As in most mean-field
approaches, this close interdependence demands that the equations 
be solved self-consistently. In particular,
the self-consistent procedure culminates with a few observables that 
fully characterize the mean-field ground state. 

With these observables
in place, one may proceed to compute the linear response of the 
mean-field ground state to an external perturbation. All dynamical 
information relevant to the excitation spectrum of the system is encoded 
in the \emph{polarization propagator} which is a function of both the 
energy and momentum transfer to the nucleus\,\cite{Fetter:1971,Dickhoff:2005}. 
The first step in the calculation of the nuclear response is the construction 
of the \emph{uncorrelated} (or mean-field) polarization propagator. That is,
\begin{align}
  \Pi_{ab}({\bf x},{\bf y};\omega) &= \sum_{0<n<F} 
   \!\overline{U}_{n}({\bf x})\Gamma^{a}
    G_{F}\Big({\bf x},{\bf y};+\omega\!+\!E_{n}^{(+)}\Big) 
    \Gamma^{b}\,U_{n}({\bf y})  \nonumber \\
    &+ \; \sum_{0<n<F} 
    \!\overline{U}_{n}({\bf y})\Gamma^{b}
    G_{F}\Big({\bf y},{\bf x};-\omega\!+\!E_{n}^{(+)}\Big) 
    \Gamma^{a}\,U_{n}({\bf x}) \,,
\label{Piab}
\end{align}
where $E_{n}^{(+)}$ and $U_{n}({\bf x})$ are single-particle
energies and Dirac wave-functions obtained from the 
self-consistent determination of the mean-field ground state,
$\Gamma^{a}$ contains information on the Lorentz and isospin 
structure of  the operator responsible for the transition, and $G_{F}$ 
is the ``Feynman'' propagator. Note that the sum is restricted to 
positive-energy states below the Fermi level. In turn,  $G_{F}$ may 
also be expressed as a sum over states by relying on its spectral 
decomposition:
\begin{equation}
  G_{F}({\bf x},{\bf y};\omega) = \sum_{n}
   \left(
     \frac{U_{n}({\bf x})\overline{U}_{n}({\bf y})}
          {\omega - E_{n}^{(+)} + i\eta} + 
     \frac{V_{n}({\bf x})\overline{V}_{n}({\bf y})}
          {\omega + E_{n}^{(-)} - i\eta}  
   \right) \,,
 \label{GFeyn}
\end{equation}
where now $E_{n}^{(-)}$ and $V_{n}({\bf x})$ represent single-particle
energies and Dirac wave-functions associated with the negative-energy
part of the spectrum; recall that in the relativistic formalism the positive
energy part of the spectrum by itself is not complete. However, note that
now the spectral sum is unrestricted, as it involves bound and continuum 
states of positive and negative energy. In practice, one avoids carrying 
out this infinite sum by invoking a \emph{non-spectral} representation 
of the Feynman propagator. This involves solving the following 
inhomogeneous differential equation for the Green's function in the 
mean-field approximation. That is,
\begin{equation}
  \Big(\omega\gammai{0}+i{\bm\gamma}\!\cdot\!{\bf\nabla}\!-\!M
  \!-\!\Sigma_{\rm MF}({\bf x})\Big)G_{F}({\bf x},{\bf y};\omega)=
  \delta({\bf x}-{\bf y}) \,,
 \label{GFeynEq}
\end{equation}
where $\gammai{0}$ and 
${\bm\gamma}\!=\!(\gammai{1},\gammai{2},\gammai{3})$
are Dirac gamma matrices, and $\Sigma_{\rm MF}$ is the 
\emph{same exact} mean-field potential obtained from the 
self-consistent solution of the ground-state 
problem\,\cite{Piekarewicz:2013bea}.

\begin{figure}[ht]
 \vspace{-0.1cm}
 \begin{center}
  \includegraphics[width=0.40\linewidth,angle=0]{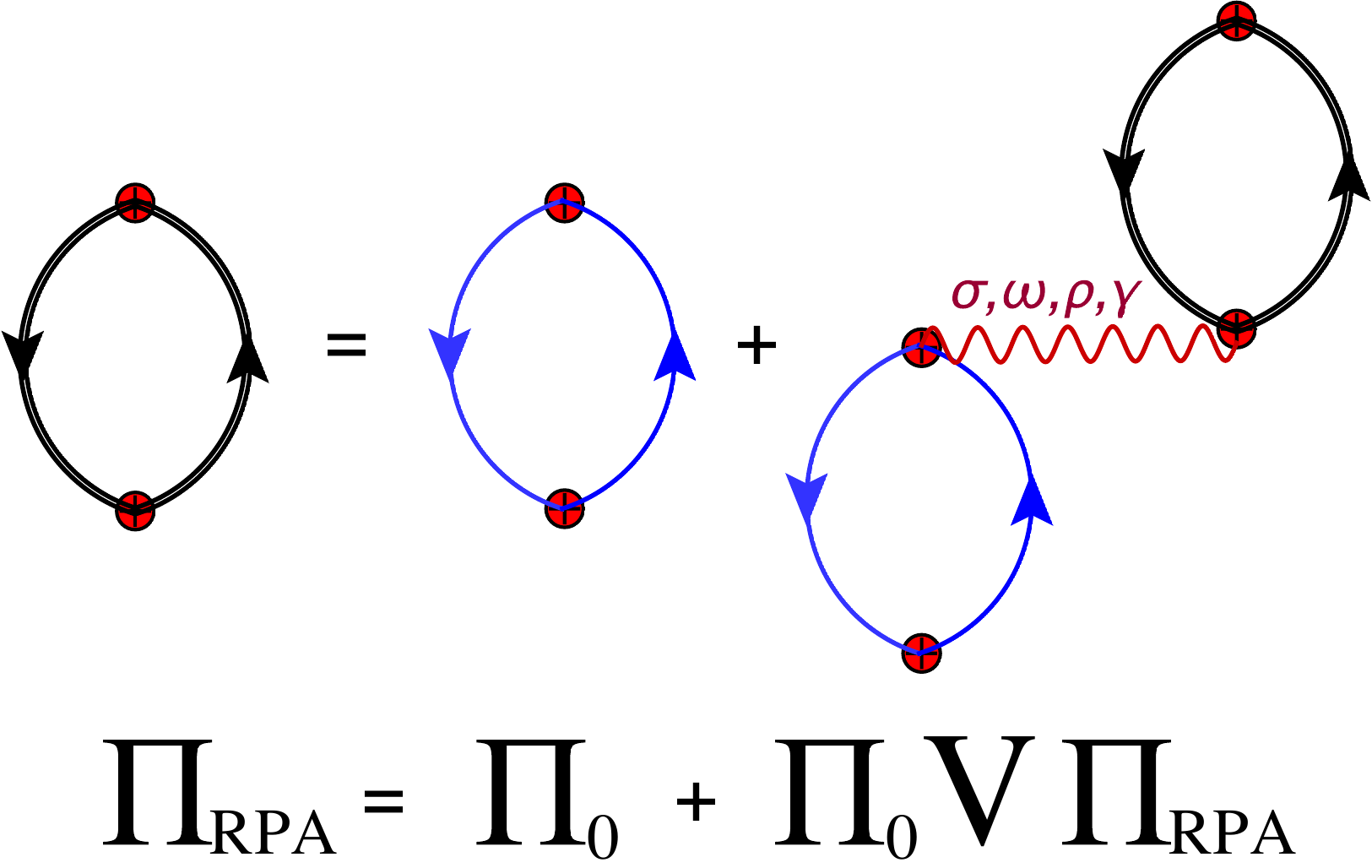}
  \caption{(Color online) Diagrammatic representation of the RPA  
  equations. The ring diagram with the thick black lines represents the fully 
  correlated RPA polarization while the one depicted with the thin blue lines is 
  the uncorrelated mean-field polarization. The residual interaction denoted with 
  the red wavy line must be identical to the one used to generate the mean-field 
  ground state.}
 \label{Fig1}
 \end{center} 
 \vspace{-0.25cm}
\end{figure}

The uncorrelated mean-field polarization together with the residual 
interaction constitute the two main building blocks of the RPA 
polarization depicted in Fig.\ref{Fig1}. By iterating the uncorrelated 
polarization to all orders, coherence is build through the mixing of all 
particle-hole excitations of the same spin and parity. If many particle-hole 
pairs get mixed, then the resulting RPA response displays strong collective 
behavior that manifests itself in the appearance of one ``giant resonance" 
that often exhausts the classical sum rule\,\cite{Harakeh:2001}. 
If the residual interaction is attractive in the channel of interest (as in the 
case of the isoscalar monopole response) then the RPA distribution of 
strength is softened and enhanced relative to the uncorrelated response. If 
instead the residual interaction is repulsive (as in the case of the isovector 
dipole response) then the RPA response is hardened and quenched. For 
further details see Refs.\,\cite{Piekarewicz:2013bea,Piekarewicz:2014bva}.

Given that the distribution of isoscalar monopole strength is the main
focus of this paper, we now briefly describe how to extract the response
from the RPA polarization. To excite the appropriate 
$J^{\pi}\!=\!0^{+},T\!=\!0$ channel, it is sufficient to use for the transition 
operator given in Eq.\,(\ref{Piab}) the time-like component of the Dirac 
gamma matrices, namely, $\Gamma^{a}\!=\!\Gamma^{b}\!=\!\gammai{0}$. 
That is, 
\begin{equation}
  S(q,\omega;E0)\!=\!-
  \frac{1}{\pi} {\rm Im}\Big(\Pi_{00}^{\rm RPA}({\bf q},{\bf q};\omega)\Big) \,.
  \label{SRPA}
\end{equation}
The above expression is still a function of both the energy and momentum 
transfer to the nucleus. In particular, ``peaks" in the energy transfer are
associated to the nuclear excitations and the area under the peaks are 
proportional to the transition form factor at the given momentum transfer.
Since inelastic $\alpha$ scattering at forward angles is the experimental
method of choice in isolating the isoscalar monopole strength from the
overall cross section, we compute $S(q,\omega;E0)$ in the long 
wavelength limit. That is,
\begin{equation}
  R(\omega;E0) = 
  \lim_{q\rightarrow 0} \left(\frac{36}{q^{4}}\right) S(q,\omega;E0) \,. 
  \label{RGMR}
\end{equation}
Often one is interested in computing moments of the strength distribution 
to extract centroid energies of the GMR. The moments of the distribution 
are defined as suitable energy weighted sums:
\begin{equation}    
  m_{n}(E0) \equiv \int_{0}^{\infty}\!\omega^{n} R(\omega;E0)\, d\omega \,.
 \label{GMRMoments}
\end{equation}
Here we concentrate on the energy weighted $m_{1}$, energy 
unweighted $m_{0}$, and inverse energy weighted $m_{-1}$ 
sums\,\cite{Harakeh:2001}. 

A few comments that highlight the power of the non-spectral approach 
are in order. First, concerning the uncorrelated mean-field polarization 
depicted by the thin blue ``bubble" in Fig.\ref{Fig1}, the non-spectral 
framework adopted here is immune to most of the features that hinder 
the spectral approach, such as an energy cutoff and the discretization 
of the continuum. Indeed, some of these artifacts were responsible for 
the faulty interpretation of the structure of the low-energy monopole 
strength observed in ${}^{68}$Ni\,\cite{Vandebrouck:2014}. Second, 
besides avoiding any reliance on artificial cutoffs and truncations, the 
non-spectral approach has the added benefit that both the positive- 
and negative-energy continua are treated exactly. Finally, in the context 
of the RPA response, it is important to underscore that in the interest of 
consistency, both the mean-field potential $\Sigma_{\rm MF}$ appearing 
in Eq.\,(\ref{GFeynEq}) as well as the residual particle hole interaction $V$ 
depicted by the red wavy line in Fig.\ref{Fig1}, must be fully consistent with 
the interaction used to generate the mean-field ground state. It is only in 
this manner that one can guarantee both the conservation of the vector
current and the decoupling of the center-of-mass motion from the physical
response\,\cite{Thouless:1961,Dawson:1990wp,Ring:2004}. 

We close this section by briefly addressing the emergence of low
energy strength in neutron-rich nuclei and its connections to
fundamental parameters of the equation of state (EOS). Evidence 
of a soft dipole mode in exotic nuclei has generated considerable 
excitement as both a novel mode of excitation and as a possible 
constraint on 
the EOS\,\cite{Tsoneva:2003gv,Piekarewicz:2006ip,
Tsoneva:2007fk,Klimkiewicz:2007zz,Carbone:2010az,
Papakonstantinou:2013gza,Savran:2013bha}. One of the goals
of the present study is to investigate how does the distribution
of isoscalar monopole strength in the calcium isotopes---particularly 
the appearance of a soft monopole mode---scales with mass number.
To quantify the sensitivity of the bulk parameters of infinite nuclear matter 
to the the isoscalar monopole response we introduce the energy per particle 
of asymmetric nuclear matter at zero temperature: 
\begin{equation}
  E/A(\rho,\alpha) - M \equiv {\cal E}(\rho,\alpha)
                          = {\cal E}_{\rm SNM}(\rho)
                          + \alpha^{2}{\cal S}(\rho)
                          + {\mathcal O}(\alpha^{4}) \,,
 \label{EOSa}
\end {equation}
where $\rho\!=\!\rho_{n}\!+\!\rho_{p}$ is the nuclear density, 
${\cal E}_{\rm SNM}$ is the energy per particle of symmetric 
nuclear matter, ${\cal S}$ is the symmetry energy, and $\alpha$
the neutron-proton asymmetry. Now expanding the energy per 
particle around saturation density ($\rhoz$) one obtains
\begin{equation}
  {\cal E}(\rho,\alpha) = \Big(\epsz +
  \frac{1}{2}K_{0}x^{2}+\frac{1}{6}Q_{0}x^{3} +\ldots\Big) + 
  \alpha^{2}\Big(J + Lx + \frac{1}{2}K_{\rm sym}x^{2}
                 +\frac{1}{6}Q_{\rm sym}x^{3} +\ldots\Big) 
                 + {\mathcal O}(\alpha^{4}) \,,
  \label{EOSb}
\end {equation}
where $x\!=\!(\rho\!-\!\rhoz)/3\rhoz$ quantifies the deviation of 
the density from its  value at saturation---and $\epsz$, 
$K_{0}$, and $Q_{0}$ denote the binding energy per nucleon, 
curvature (i.e., incompressibility), and skewness parameter of 
symmetric nuclear matter. In turn, $J$, $K_{\rm sym}$, and 
$Q_{\rm sym}$ represent the corresponding quantities for the 
symmetry energy. However, unlike symmetric nuclear matter 
which saturates, the symmetry pressure---or equivalently the 
slope of the symmetry energy $L$---does not vanish. In particular, 
$L$ induces changes in the saturation density and incompressibility 
coefficient of \emph{asymmetric neutron-rich 
matter}\,\cite{Piekarewicz:2008nh} that are given by
\begin{subequations}
 \begin{align}
  & \rhoz(\alpha)=\rhoz + \rho_{\tau}\alpha^{2}\,;
   \hspace{5pt}{\rm with} \hspace{5pt} 
   \rho_{\tau}\equiv-3\rhoz\frac{L}{K_{0}},
   \label{RhoTau} \\
  & K_{0}(\alpha)=K_{0}+K_{\tau}\alpha^{2} \,;
   \hspace{5pt}{\rm with} \hspace{5pt} 
   K_{\tau}\equiv K_{\rm sym}-6L-\frac{Q_{0}}{K_{0}}L.
 \label{KTau}
 \end{align}  
 \label{Taus}
\end{subequations}
The above expression for $K_{\tau}$ suggests that measurements 
of the isotopic dependence of the giant monopole resonance in 
exotic nuclei with a very large neutron excess---such as the calcium
isotopes---may place important constraints on the density dependence 
of the symmetry energy\,\cite{Garg:2006vc,Piekarewicz:2007us}.
Important first steps along this  direction have been already taken 
by Garg and collaborators\,\cite{Li:2007bp,Li:2010kfa,Patel:2012zd}.

\section{Results}
\label{Results}

We start this section by computing the distribution of isoscalar 
monopole strength for three relativistic mean-field models: 
(a) NL3\,\cite{Lalazissis:1996rd,Lalazissis:1999}, 
FSUGold\,\cite{Todd-Rutel:2005fa}, and 
FSUGarnet\,\cite{Chen:2014mza}. Among these three, FSUGarnet 
is a recently calibrated parametrization that has been fitted to the 
ground-state properties of magic and semi-magic nuclei, a few 
giant monopole energies, and well established properties of neutron 
stars\,\cite{Chen:2014sca}. In particular, FSUGarnet predicts that
the isotopic chain in oxygen can be made to terminate at ${}^{24}$O, 
as it has been observed experimentally\,\cite{Thoennessen:2004}. 
The parameters for the three relativistic models adopted in this work 
are listed in Table\,\ref{Table1} in terms of the Lagrangian density 
defined in Ref.\,\cite{Chen:2014sca}. 

\begin{table}[h]
\begin{tabular}{|l||c|c|c|c|c|c|c|c|c|c|}
 \hline\rule{0pt}{2.25ex}   
 \!\!Model & $m_{\rm s}$  & $m_{\rm v}$  & $m_{\rho}$  
       & $g_{\rm s}^2$ & $g_{\rm v}^2$ & $g_{\rho}^2$
       & $\kappa$ & $\lambda$ & $\zeta$ & $\Lambda_{\rm v}$\\
 \hline
 \hline
 NL3                   & 508.194 & 782.501 & 763.000 & 104.3871 & 165.5854 &  79.6000 
                           & 3.8599  & $-$0.015905 & 0.000000 & 0.000000 \\
 FSUGold            & 491.500 & 782.500 & 763.000 & 112.1996 & 204.5469 & 138.4701 
                            & 1.4203  & $+$0.023762 & 0.060000 & 0.030000 \\
 FSUGarnet       & 496.939 & 782.500 & 763.000 &  110.3492 & 187.6947 & 192.9274
                          & 3.2602  & $-$0.003551 & 0.023500 & 0.043377 \\
\hline
\end{tabular}
\caption{Parameter sets for the three models adopted in the text. All meson masses
  ($m_{\rm s}$, $m_{\rm v}$,  and $m_{\rho}$) as well as $\kappa$ are given in MeV. 
  The nucleon mass has been fixed in all models at $M\!=\!939$~MeV. See 
  Ref.\,\cite{Chen:2014sca} for a definition of the model parameters.}
\label{Table1}
\end{table}

Using the model parameters listed in Table\,\ref{Table1}, one can 
then proceed to make predictions for a variety of bulk properties 
of  infinite nuclear matter as defined in Eqs.\,(\ref{EOSb}) 
and\,(\ref{Taus}). We display these set of bulk properties in Table\,\ref{Table2} 
alongside quantities denoted as $K_{40}$, $K_{48}$, and $K_{60}$ 
that represent the incompressibility coefficient of asymmetric matter 
having the same neutron excess as ${}^{40}$Ca($\alpha\!=\!0$), 
${}^{48}$Ca($\alpha\!\simeq\!0.167$), and 
${}^{60}$Ca($\alpha\!\simeq\!0.333$), respectively. Although NL3 
predicts an incompressibility coefficient $K_{0}$ that is considerably 
larger than the other two models, NL3 becomes the
softest of the three models for infinite nuclear matter with a 
neutron excess equal to that of ${}^{60}$Ca. This rapid softening
of the incompressibility coefficient is attributed to NL3's stiff symmetry 
energy ($L\!\simeq\!118$\,MeV) which in turn is responsible for 
generating a large and negative value for $K_{\tau}$. Hence, 
exploring the isotopic dependence of the ISGMR over an isotopic 
chain that includes exotic nuclei with a large neutron excess may 
become instrumental in constraining the density dependence of the 
symmetry energy.

\begin{table}[h]
\begin{tabular}{|l||c|c|c|r|r|r|r||r|r|r|r|}
 \hline\rule{0pt}{2.25ex}   
 \!\!Model & $\rhoz$ & $\epsz$ 
           & $K_{0}$& $\hfill Q_{0}\hfill$& $\hfill J\hfill$& $\hfill L\hfill$& 
           $\hfill K_{\rm sym}\hfill$& $\hfill K_{\tau} \hfill$& $\hfill K_{40} \hfill$
           & $\hfill K_{48} \hfill$ & $\hfill K_{60} \hfill$\\
 \hline
 \hline
 NL3              &  0.148  & $-$16.24 & 271.5 & 209.5 & 37.29 & 118.2 &        100.9 &  $-$699.4 & 271.5 & 252.1 & 193.8\\ 
 FSUGold      &  0.148  & $-$16.30 & 230.0 & $-$522.7 & 32.59 & 60.5 & $-$51.3 & $-$276.9 & 230.0  & 222.3 & 199.3\\
 FSUGarnet  & 0.153  & $-$16.23 & 229.5 & 4.5 & 30.92 & 51.0  &         59.5 & $-$247.3 & 229.5  & 222.7 &202.1\\
\hline
\end{tabular}
\caption{Bulk parameters characterizing the behavior of infinite nuclear matter 
             at saturation density as defined in Eqs.\,(\ref{EOSb}) and\,(\ref{Taus}).
             Here $K_{40}$, $K_{48}$, and $K_{60}$ represent the incompressibility 
             coefficient of asymmetric matter having the same neutron excess as 
             ${}^{40}$Ca, ${}^{48}$Ca, and ${}^{60}$Ni, respectively. All quantities 
             are given in MeV except for $\rho_{{}_{0}}$ which is given in 
             ${\rm fm}^{-3}$.}
\label{Table2}
\end{table}

\begin{figure}[ht]
\centering
 \includegraphics[width=.5\linewidth]{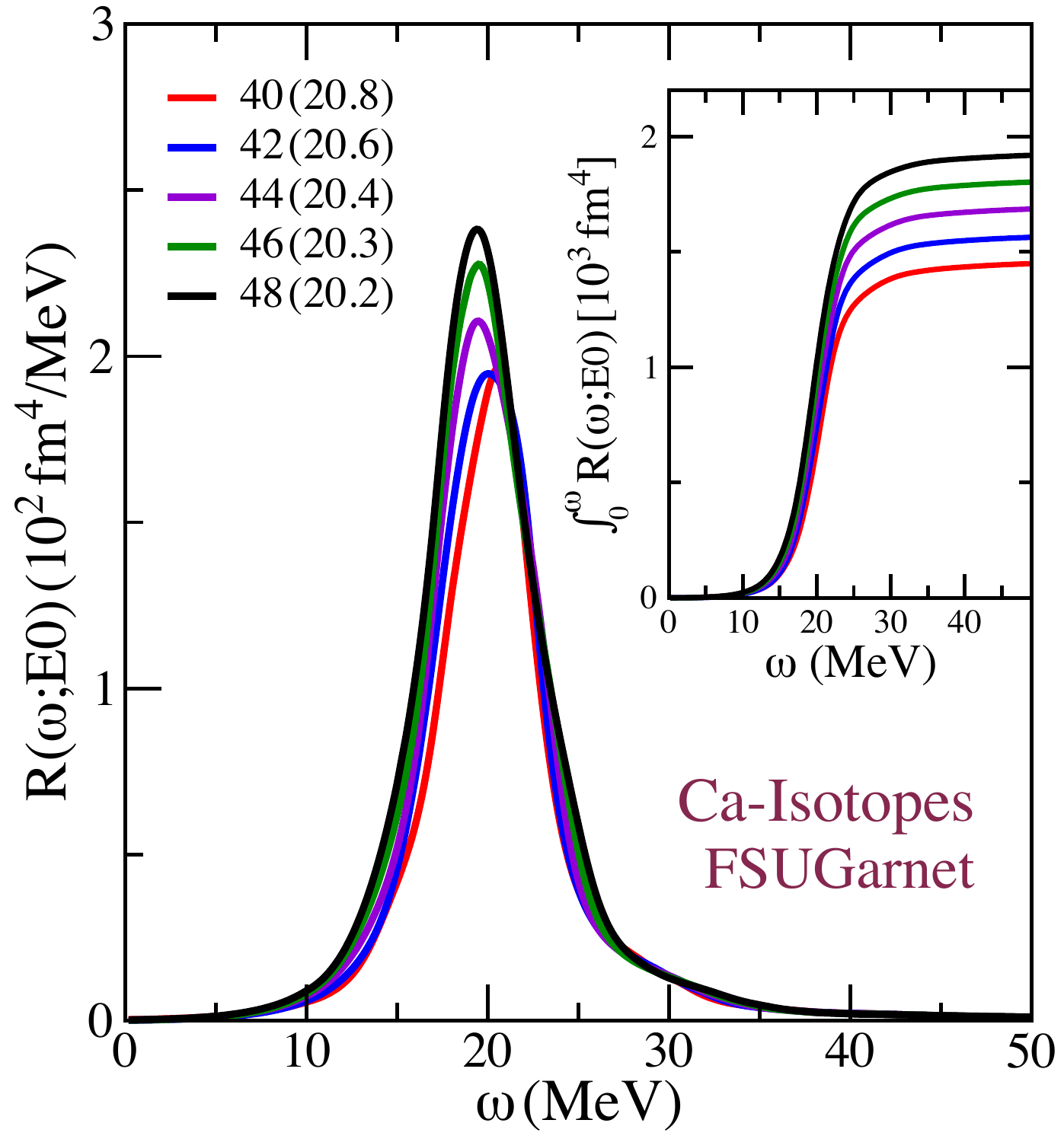}
\caption{(Color online) (a) Distribution of isoscalar monopole strength 
for the neutron-even calcium isotopes from ${}^{40}$Ca up to ${}^{48}$Ca 
as predicted by a relativistic RPA calculation using the FSUGarnet 
parametrization\,\cite{Chen:2014mza}. The numbers in parenthesis 
represent the centroid energies defined as $E_{\rm cen}\!=\!m_{1}/m_{0}$.
The inset displays the integrated strength, or ``running sum", with the 
value at large excitation energy equal to the unweighted energy sum 
$m_{0}$.}
\label{Fig2}
\end{figure}

The distribution of isoscalar monopole strength for all stable even-even calcium 
isotopes from ${}^{40}$Ca to ${}^{48}$Ca is displayed in Fig.\,\ref{Fig2} as 
predicted by the new FSUGarnet parametrization with the centroid energies 
$m_{1}/m_{0}$ shown in parenthesis; see Eq.\,(\ref{GMRMoments}) and
Table\,\ref{Table3}. Although an enhancement of the response is evident as the 
f${}^{\,7/2}$ neutron orbital is progressively being filled (see inset in the figure) no 
appreciable softening is observed. This is somehow surprising given that as the 
f${}^{\,7/2}$ orbital is being filled and the neutron-skin thickness steadily increases,
the appearance of a soft monopole mode involving neutron-skin oscillations may 
have been natural. Yet, no low-energy monopole strength is detected. Values 
for the neutron-skin thickness of the calcium isotopes alongside other ground 
state properties are listed in Table\,\ref{Table4}.

\begin{table}[h]
  \begin{tabular}{|c||c|c||c|c||c|c||c|c|}
   \cline{2-9}
   \hline\rule{0pt}{2.5ex} 
   & \multicolumn{2}{c||}{NL3\cite{Lalazissis:1996rd,Lalazissis:1999}} &
       \multicolumn{2}{c||}{FSUGold\cite{Todd-Rutel:2005fa}} &
       \multicolumn{2}{c||}{FSUGarnet\cite{Chen:2014mza}} &     
       \multicolumn{2}{c|}{Experiment\cite{Youngblood:2001mq,Lui:2011zz}} \\  
    \cline{2-9}
   \hline\rule{0pt}{2.5ex} 
    Isotope & $m_{1}/m_{0}$ & $\sqrt{m_{1}/m_{-1}}$  &
       $m_{1}/m_{0}$ & $\sqrt{m_{1}/m_{-1}}$ &
       $m_{1}/m_{0}$ & $\sqrt{m_{1}/m_{-1}}$ &
       $m_{1}/m_{0}$ & $\sqrt{m_{1}/m_{-1}}$ \\
    \hline\rule{0pt}{2.5ex}
    \!\!${}^{40}$Ca & 22.29 & 21.52 & 21.53 & 20.76 & 20.81 & 20.09  
                            & $19.2\pm0.4$ & $18.3\pm0.4$\\
         ${}^{42}$Ca & 22.04	& 21.28 & 21.39 & 20.74 & 20.64 & 20.02  & & \\
         ${}^{44}$Ca & 21.81	& 21.10 & 21.16 & 20.47 & 20.44 & 19.81 & & \\
         ${}^{46}$Ca & 21.56	& 20.89 & 21.00 & 20.34 & 20.31 & 19.69 &  & \\
         ${}^{48}$Ca & 21.27	& 20.58 & 20.85 & 20.20 & 20.22 & 19.57 
                            & $19.88^{+0.14}_{-0.18}$ & $19.04^{+0.11}_{-0.14}$ \\
    \hline
 \end{tabular}
\caption{Predictions for the centroid energies (all in MeV) for all stable
even-even calcium isotopes obtained through two different moment-ratios 
of the isoscalar monopole response displayed Fig.\,\ref{Fig2} integrated up 
to a maximum value of $\omega_{\rm max}\!=\!50$\,MeV.}
 \label{Table3}
\end{table}

The predicted trend displayed in Fig.\,\ref{Fig2}, however, is consistent with 
the expectation of a softening of the response with increasing mass 
number\,\cite{Harakeh:2001}. Hence, experimental evidence in favor of a 
centroid energy that is actually \emph{lower} in ${}^{40}$Ca than it is in 
${}^{48}$Ca\,\cite{Youngblood:2001mq,Lui:2011zz} 
has baffled theoretical explanations; see Ref.\,\cite{Anders:2013rea} and references 
contained therein. As illustrated in Table\,\ref{Table3}, we too have no explanation 
for this apparent anomaly. Mapping the experimental distribution of strength 
in the intermediate region, from ${}^{42}$Ca to ${}^{46}$Ca, is encouraged 
as it may ultimately proved vital in solving the puzzle.

\begin{figure}[ht]
\centering
 \includegraphics[width=.5\linewidth]{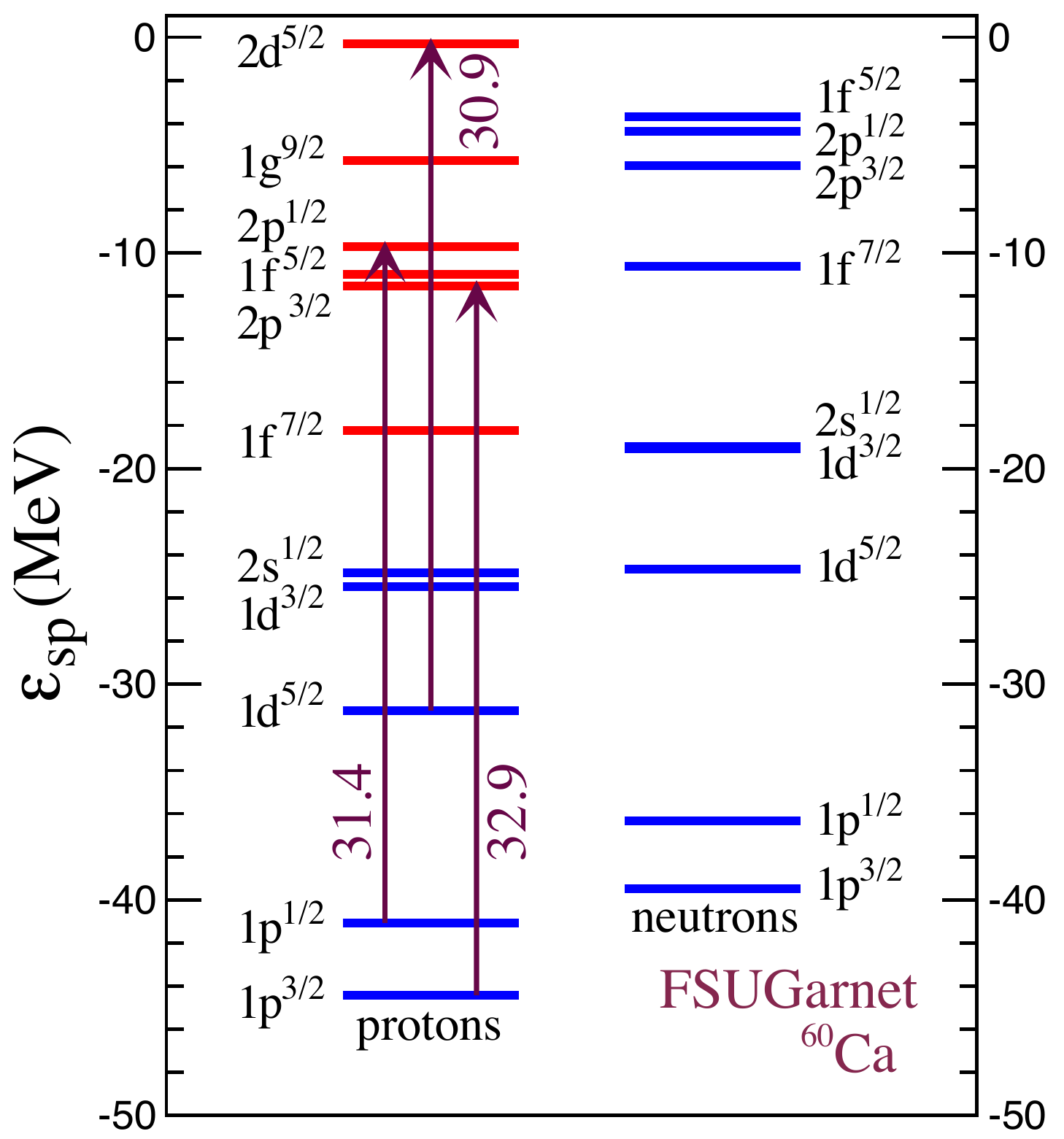}
\caption{(Color online) Single-particle spectrum for ${}^{60}$Ca
as predicted by the relativistic FSUGarnet parametrization. The
blue(red) lines denote occupied(empty) orbitals and the arrows 
indicate discrete transitions into bound states.}
\label{Fig3}
\end{figure}

\begin{table}[h]
\begin{tabular}{|c||c|c|c|c||c|c|c|c||c|c|c|c|}
   \hline\rule{0pt}{2.5ex} 
   & \multicolumn{4}{c||}{NL3\cite{Lalazissis:1996rd,Lalazissis:1999}} &
       \multicolumn{4}{c||}{FSUGold\cite{Todd-Rutel:2005fa}} &
       \multicolumn{4}{c|}{FSUGarnet\cite{Chen:2014mza}} \\  
   \hline\rule{0pt}{2.5ex} 
    Isotope & $B/A$  & $R_{\rm ch}$ & $R_{n}$ & $R_{\rm skin}$ & 
                    $B/A$  & $R_{\rm ch}$ & $R_{n}$ & $R_{\rm skin}$  &
                    $B/A$  & $R_{\rm ch}$ & $R_{n}$ & $R_{\rm skin}$ \\     
     \hline\rule{0pt}{2.5ex}
    \!\!${}^{40}$Ca &  -8.543 & 3.329 & 3.377 & -0.048 & -8.539 & 3.287 & 3.338 & -0.051 & -8.531 & 3.293 & 3.345 & -0.052\\
        ${}^{42}$Ca &  -8.552 & 3.412 & 3.374  & 0.038 & -8.538 & 3.372 & 3.342  & 0.030  & -8.539 & 3.369 & 3.347 & 0.022\\      
	${}^{44}$Ca &  -8.572 & 3.484 & 3.373 & 0.111 & -8.546 & 3.445 & 3.348  & 0.097  & -8.554  & 3.430 & 3.350 & 0.081\\
	${}^{46}$Ca &  -8.603 & 3.548 & 3.375 & 0.172 & -8.561 & 3.508 & 3.356 & 0.152  & -8.578  & 3.482 & 3.353 & 0.128\\
	${}^{48}$Ca &  -8.641 & 3.605 & 3.379 & 0.226 & -8.585 & 3.563 & 3.366 & 0.197  & -8.613  & 3.524 & 3.357 & 0.167\\
	${}^{50}$Ca &  -8.489 & 3.732 & 3.398 & 0.334 & -8.421 & 3.693 & 3.388 & 0.305  & -8.448  & 3.651 & 3.384 & 0.267\\
	${}^{52}$Ca &  -8.367 & 3.839 & 3.416 & 0.423 & -8.284 & 3.802 & 3.411 & 0.391  & -8.313  & 3.755 & 3.412 & 0.343\\
	${}^{54}$Ca &  -8.204 & 3.962 & 3.438 & 0.524 & -8.115 & 3.921 & 3.435 & 0.486  &-8.135  & 3.874 & 3.440 & 0.434\\
	${}^{56}$Ca &  -8.038 & 4.035 & 3.476 & 0.559 & -7.937 & 3.982 & 3.471 & 0.512  &-7.923  & 3.943 & 3.482 & 0.461\\
	${}^{58}$Ca &  -7.903 & 4.100 & 3.514 & 0.586 & -7.789 & 4.036 & 3.506 & 0.530  & -7.746  & 4.002 & 3.523 & 0.478\\
	${}^{60}$Ca &  -7.793 & 4.159 & 3.551 & 0.608 & -7.667 & 4.084 & 3.541 & 0.543  & -7.600  & 4.051 & 3.563 & 0.488\\
    \hline
 \end{tabular}
\caption{Predictions for the binding energy per nucleon, charge radius, 
neutron radius, and neutron-skin thickness for all even-even $N\!\ge\!20$ 
calcium isotopes. Binding energies are given in MeV and radii in fm.}
 \label{Table4}
\end{table}

\begin{figure}[ht]
\vspace{-0.05in}
\includegraphics[width=.4\linewidth]{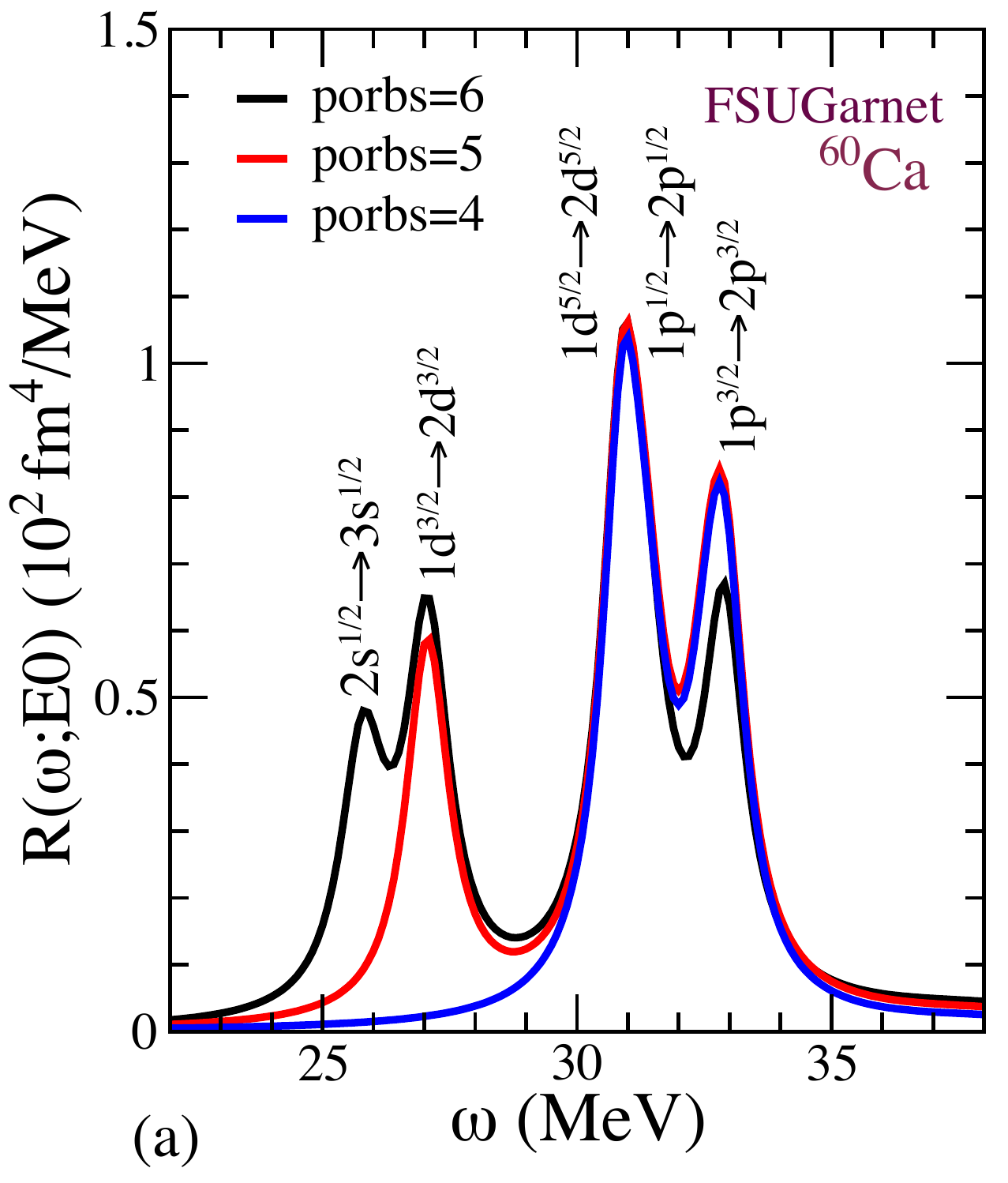}
 \hspace{0.2cm}
\includegraphics[width=.4\linewidth]{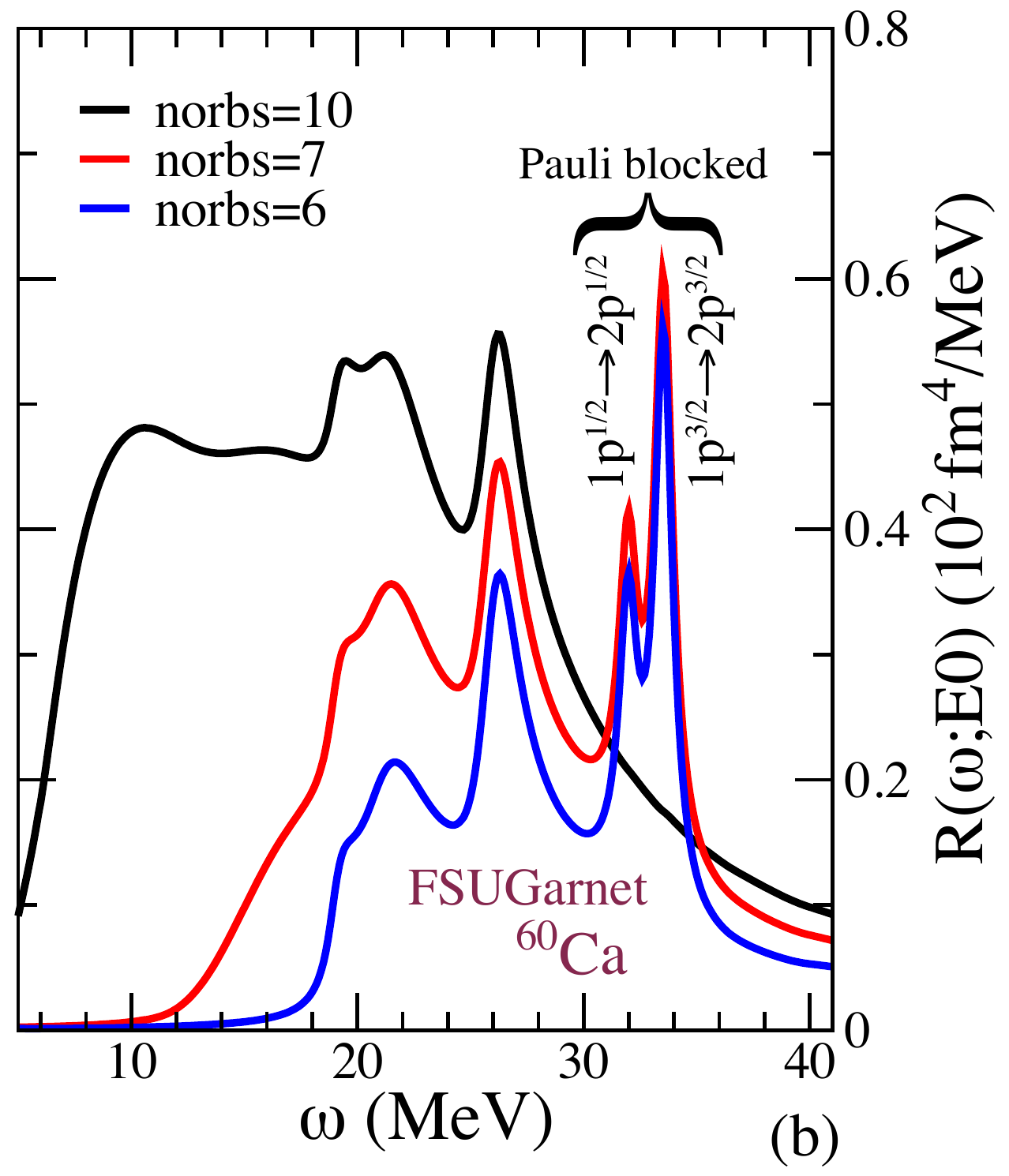}
\caption{(Color online) Distribution of ``uncorrelated'' isoscalar  monopole 
strength for ${}^{60}$Ca as predicted by the FSUGarnet 
parametrization\,\cite{Chen:2014mza}. The left-hand panel (a) describes
proton particle-hole excitations while the right-hand panel (b) depicts the
corresponding neutron excitations. Transitions into bound states are easily 
identified both in this figure as well as in Fig.\,\ref{Fig3}. See text for further 
details.}
\label{Fig4}
\end{figure}

\begin{figure}[ht]
\vspace{-0.05in}
\includegraphics[height=8.1cm]{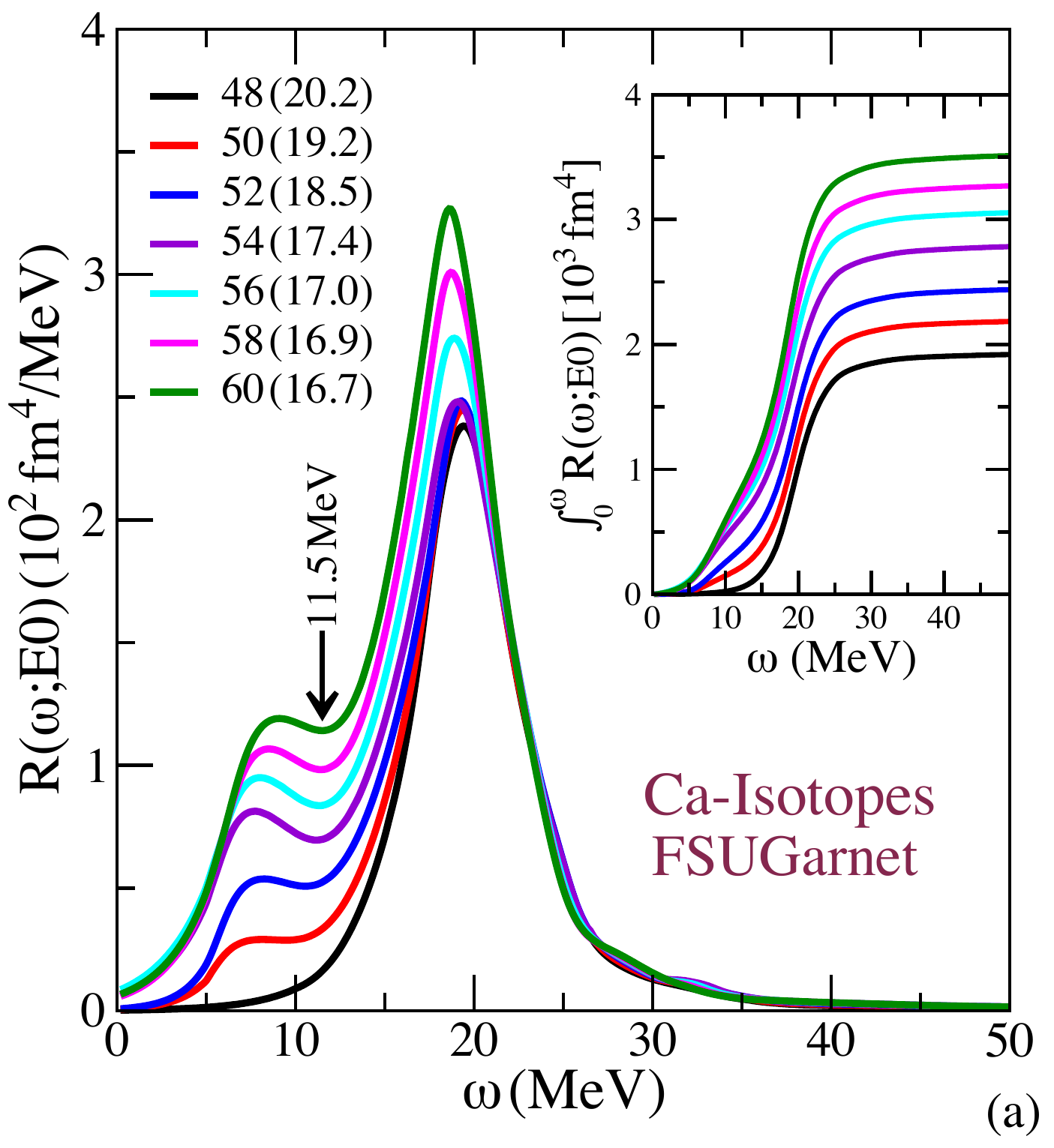}
 \hspace{0.2cm}
\includegraphics[height=8cm]{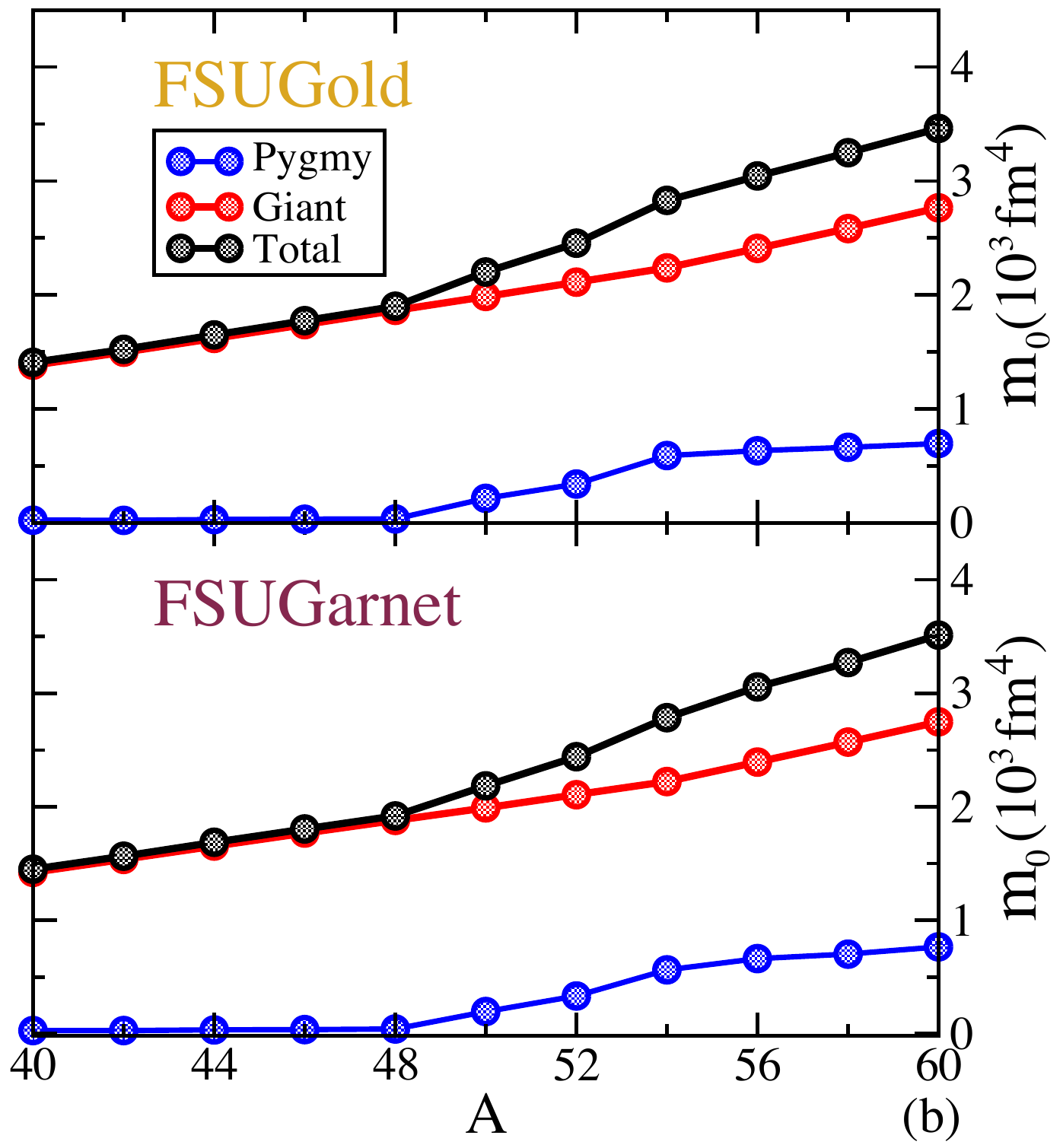}
\caption{(Color online) (a) Distribution of isoscalar monopole strength 
for the neutron-even calcium isotopes from ${}^{48}$Ca to ${}^{60}$Ca 
as predicted by a relativistic RPA calculation using the FSUGarnet 
parametrization\,\cite{Chen:2014mza}. The numbers in parenthesis 
represent the centroid energies defined as $E_{\rm cen}\!=\!m_{1}/m_{0}$.
The inset displays the integrated strength, or ``running sum", with the value 
at large excitation energy equal to the unweighted energy sum $m_{0}$. (b) 
Contribution from the low-energy (``Pygmy") and high-energy (``Giant'') 
resonance region to the total unweighted energy sum $m_{0}$ for two 
relativistic parametrizations: 
FSUGold\,\cite{Todd-Rutel:2005fa} and FSUGarnet.}
\label{Fig5}
\end{figure}

While no evidence of low-energy monopole strength was found in the stable calcium
isotopes as the 1f${}^{\,7/2}$ neutron orbital was being filled, the situation changes 
dramatically with the occupation of the weakly-bound 2p and 1f${}^{\,5/2}$orbitals. In
an effort to elucidate the nature of the low-energy monopole strength observed in the 
unstable calcium isotopes [see Fig.\ref{Fig5}(a)] it is convenient to start by displaying
in Fig.\,\ref{Fig3} the single-particle spectrum of ${}^{60}$Ca as predicted by 
FSUGarnet; the arrows are used to indicate three prominent discrete proton excitations. 
Given that information about the mean-field excitation spectrum is fully contained in the 
uncorrelated response, the three discrete proton excitations in the $\omega\!\gtrsim\!30$\,MeV 
region are clearly discernible in Fig.\,\ref{Fig4}(a) (the d${}^{5/2}$ and p${}^{1/2}$ 
excitations can not be individually resolved in the figure). Besides these discrete proton 
excitations, two additional peaks are prominent in the 25-28 MeV region that can not 
be inferred from the single-particle spectrum, as these involve transitions into the continuum. 
To illuminate the structure of these additional transitions, we artificially remove the two 
valence 2s${}^{1/2}$ and 1d${}^{3/2}$ proton orbitals, both having a binding energy of 
about 25\, MeV (see  Fig.\,\ref{Fig3}). In particular, the curve denoted by ${\rm porbs}\!=\!5$ 
(red line) was obtained by removing the 2s${}^{1/2}$ orbital. The curve clearly shows 
how all the strength below $\sim\!25$\,MeV practically disappears. This effect is further 
accentuated by removing both (2s${}^{1/2}$ and 1d${}^{3/2}$) proton orbitals, resulting 
in the blue curve denoted by ${\rm porbs}\!=\!4$. Essentially, no monopole strength 
remains below 30\, MeV.

If the role of the continuum is important in elucidating the character of the proton excitations, 
it becomes even more critical in understanding the nature of the neutron excitations---given 
that they all involve transitions into the continuum. Indeed, in Fig.\,\ref{Fig4}(b) we observe
a significant amount of low energy monopole strength resulting from the excitation into the
continuum of the weakly bound 2p and 1f${}^{\,5/2}$  states; see black line. As in the proton 
case, we underscore the role of these low-energy excitations by suppressing either three 
(red line) or four (blue line) transitions out of the valence 2p-1f orbitals. By doing so, 
essentially no monopole strength remains below $\omega\sim18$\,MeV. Moreover, it
is interesting to note the two emerging discrete transitions out of the 1p states 
at $\omega\!\gtrsim\!30$\,MeV. The fact that these two Pauli-blocked transitions are 
completely suppressed from the complete calculation (black line) further validates the 
non-spectral treatment of the nuclear response. Before addressing the fully correlated RPA 
response, it is important to underscore the virtues of the non-spectral approach, especially 
in the context of weakly-bound nuclei. The non-spectral Green's function approach adopted 
here allows for an exact treatment of the continuum without any reliance on artificial parameters.
Indeed, introducing an energy cutoff or the discretization of the continuum is neither required 
nor admitted. 

\begin{figure}[ht]
\vspace{-0.05in}
\includegraphics[height=8cm]{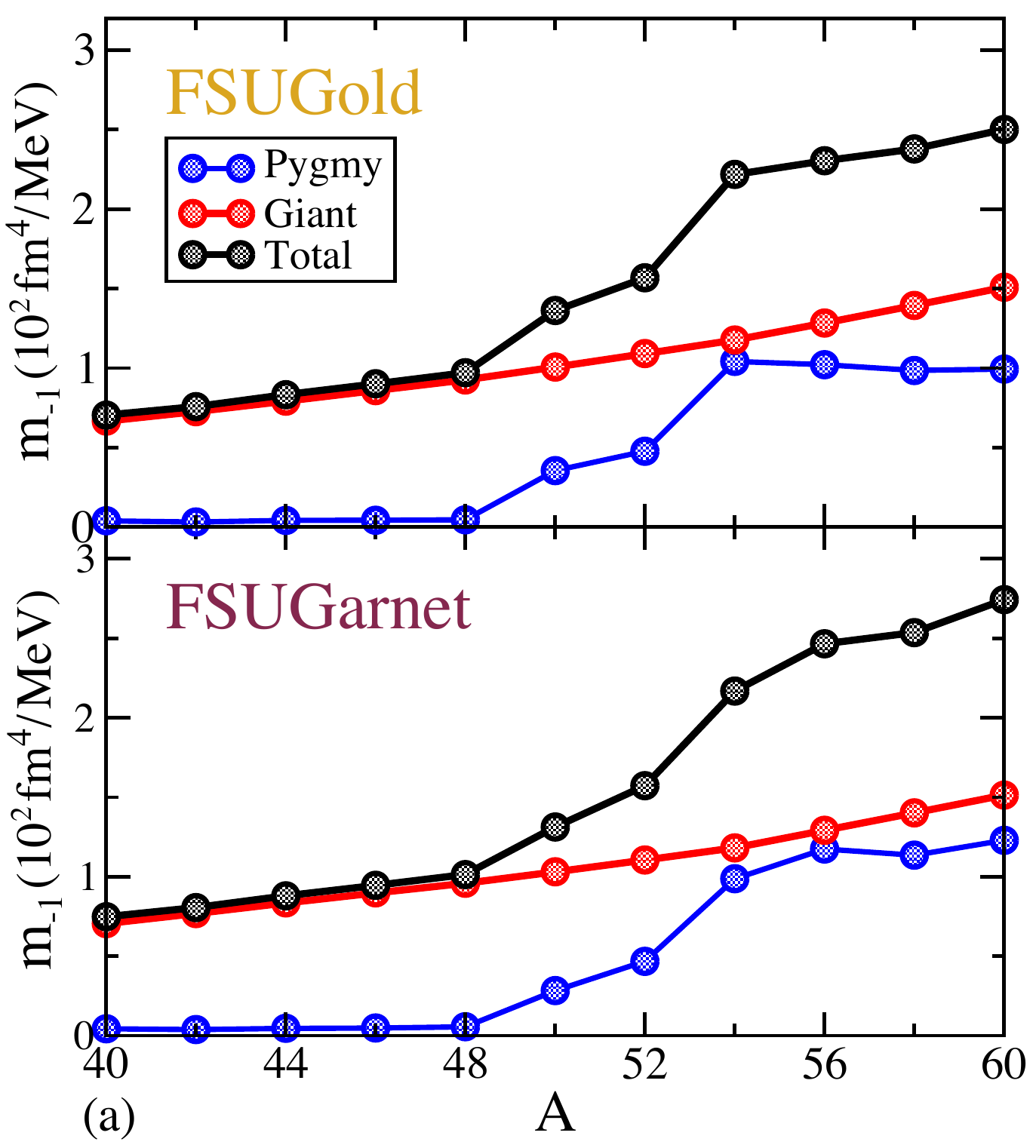}
 \hspace{0.2cm}
\includegraphics[height=8cm]{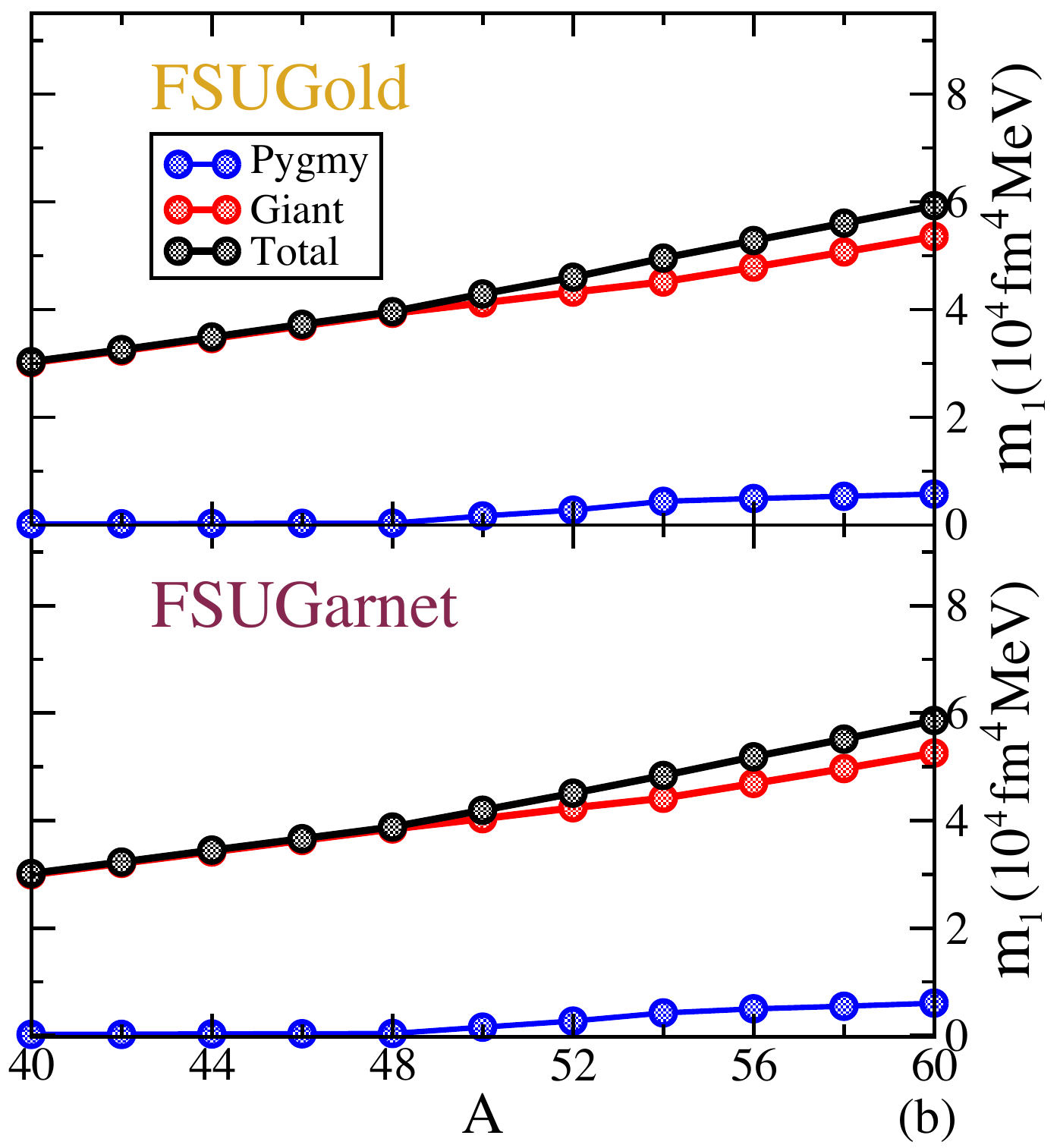}
\caption{(Color online) Contribution from the low-energy (``Pygmy") and 
high-energy (``Giant'') resonance region to the total (a) inverse energy 
weighted sum $m_{-1}$ and (b) energy weighted sum $m_{1}$, as 
predicted by two relativistic parametrizations: 
FSUGold\,\cite{Todd-Rutel:2005fa} and FSUGarnet\,\cite{Chen:2014mza}.}
\label{Fig6}
\end{figure}

\begin{table}[h]
\begin{tabular}{|c||c|c|c|c||c|c|c|c||c|c|c|c|}
   \hline\rule{0pt}{2.5ex} 
   & \multicolumn{4}{c||}{NL3\cite{Lalazissis:1996rd,Lalazissis:1999}} &
       \multicolumn{4}{c||}{FSUGold\cite{Todd-Rutel:2005fa}} &
       \multicolumn{4}{c|}{FSUGarnet\cite{Chen:2014mza}} \\  
   \hline\rule{0pt}{2.5ex} 
    Isotope & $R_{\rm skin}$  & Pygmy & Giant & Total &
                    $R_{\rm skin}$  & Pygmy & Giant & Total &
                    $R_{\rm skin}$  & Pygmy & Giant & Total \\     
     \hline\rule{0pt}{2.5ex}
 \!\!${}^{50}$Ca& 0.334  & 23.78  &   96.55 & 120.33	 & 0.305 &    35.50  & 100.61  & 136.11    & 0.267 &  28.27  & 102.99 & 131.26 \\
    ${}^{52}$Ca& 0.423  & 36.30  & 105.06 & 141.36	 & 0.391  &   47.66  & 109.01  & 156.67    & 0.343 &  46.76  & 110.45 & 157.21 \\
    ${}^{54}$Ca& 0.524  & 75.42  & 114.24 & 189.66	 & 0.486  & 104.15  & 117.62  & 221.77    & 0.434 &  98.61  & 118.15 & 216.75 \\
    ${}^{56}$Ca& 0.559  & 83.30  & 126.32 & 209.62	 & 0.512  & 102.10  & 128.28  & 230.38    & 0.461 & 117.39 & 129.09 & 246.48 \\
    ${}^{58}$Ca& 0.586  & 68.16  & 138.88 & 207.05	 & 0.530  &   98.52  & 139.36  & 237.88    & 0.478 & 113.47 & 140.13 & 253.60 \\
    ${}^{60}$Ca& 0.608  & 76.73  & 152.31 & 229.04	 & 0.543  &   99.29  & 150.83  & 250.12    & 0.488 & 122.86 & 151.37 & 274.22 \\
    \hline
 \end{tabular}
\caption{Predictions for the neutron-skin thickness, and pygmy, giant, and total contributions 
to the inverse energy weighted sum $m_{{}_{-1}}$ for all even-even calcium isotopes from
${}^{50}$Ca to ${}^{60}$Ca. An excitation energy of $\omega\!=\!11.5$\,MeVwas chosen to 
separate the pygmy from the giant resonance region; see Fig.\,\ref{Fig5}(a). The neutron-skin 
thickness is given in fm and the $m_{{}_{-1}}$ values in ${\rm fm}^{4}/{\rm MeV}$.}
 \label{Table5}
\end{table}

\begin{figure}[ht]
 \vspace{-0.1cm}
 \begin{center}
  \includegraphics[width=0.50\linewidth,angle=0]{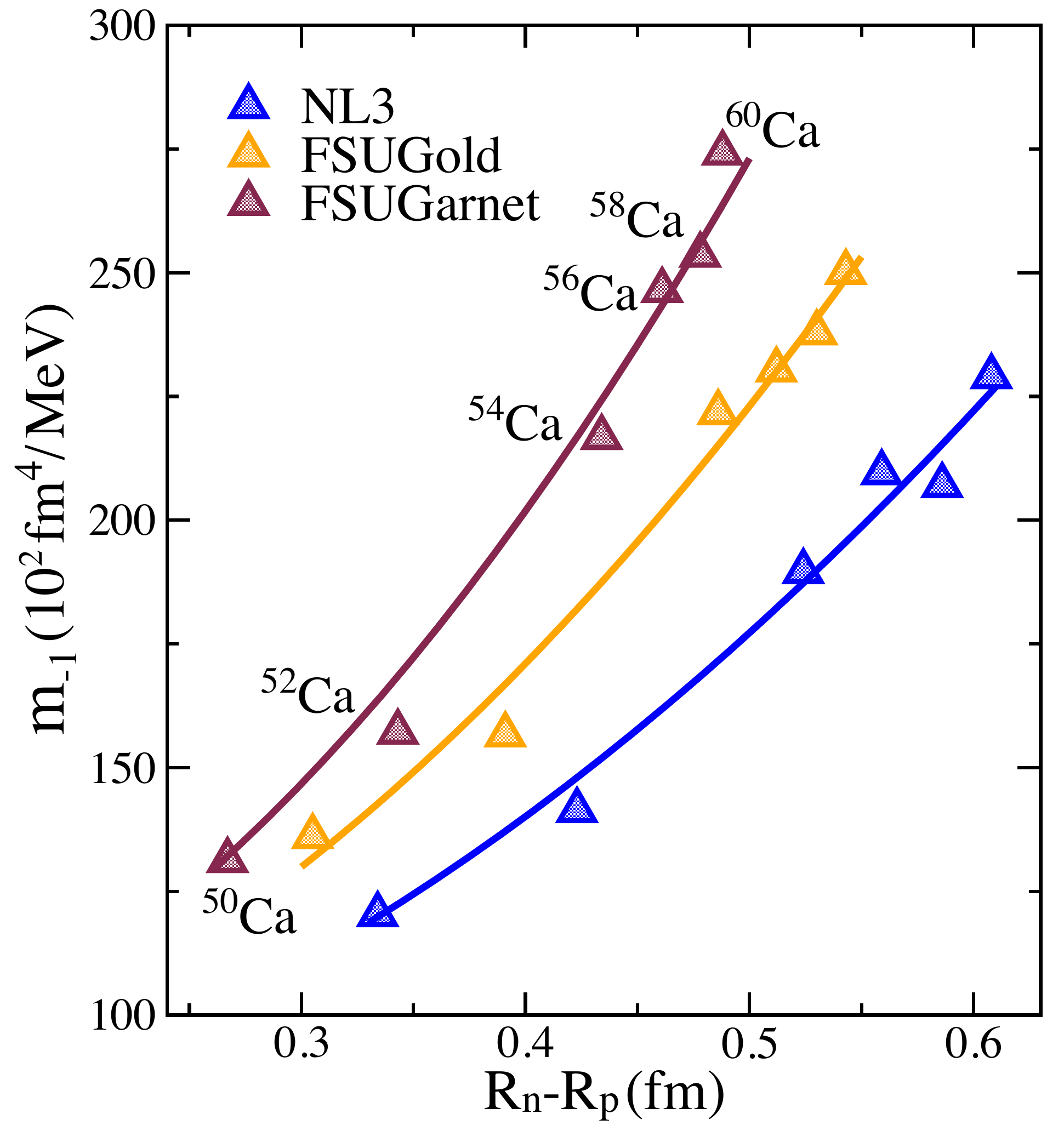}
  \caption{(Color online) Total (pygmy plus giant) inverse energy weighted
  sum as a function of the corresponding neutron skin-thickness for all
  even-even calcium isotopes from ${}^{50}$Ca to ${}^{60}$Ca. Predictions
  are displayed from all three relativistic models used in the text.} 
 \label{Fig7}
 \end{center} 
 \vspace{-0.25cm}
\end{figure}

In Fig.\,\ref{Fig1} the diagrammatic representation of the RPA equations indicates how
the correlated RPA response (solid black line) is obtained from the mean-field response
(thin blue line) and the residual particle-hole interaction (wavy red line). Given that the
RPA equations involve an iterative procedure to all orders, the singularity structure of
the polarization is often changed dramatically. In the particular case of an attractive 
residual interaction, mixing a large number of particle-hole excitations results in the 
development of a single collective peak that exhausts most of the energy weighted 
sum\,\cite{Harakeh:2001}, as is clearly evident in Fig.\,\ref{Fig2}. However, for the 
unstable neutron-rich isotopes a significant amount of low-energy strength is 
expected to emerge as the result of transitions from weakly-bound states into the 
continuum. This notion is confirmed in Fig.\,\ref{Fig5}(a) that displays the distribution 
of isoscalar monopole strength for all unstable even-even calcium isotopes from 
${}^{50}$Ca to ${}^{60}$Ca; the single-peak response of ${}^{48}$Ca is also shown
for reference. As in the case of Fig.\,\ref{Fig2}, a giant resonance peak in the 
$\sim\!17\!-\!20$\,MeV region is noticeable. Yet, a significant amount of low-energy 
monopole strength that progressively increases with mass number is also clearly 
evident. This can be quantified by plotting the running sum 
in the inset of Fig.\,\ref{Fig5}(a). In contrast to the corresponding inset displayed in 
Fig.\,\ref{Fig2}, a considerable amount of low-energy strength appears below 10 MeV.
By somehow artificially dividing the giant-monopole region from the low-energy
``Pygmy" region at an excitation energy of $\omega\!=\!11.5$\,MeV, we display in
Fig.\,\ref{Fig5}(b) both the pygmy and giant contributions to the total unweighted
energy sum $m_{0}$ as predicted by the FSUGold and FSUGarnet 
parametrizations. The figure clearly indicates the important contribution from the
low-energy region to the overall $m_{{}_{0}}$ moment. As expected, the pygmy 
contribution can be correspondingly enhanced or quenched by computing 
the inverse energy weighted sum $m_{{}_{-1}}$ or energy weighted sum 
$m_{{}_{1}}$ as shown in Figs.\,\ref{Fig6}(a) and (b), respectively.  In particular,
we observe nearly equal contributions from the pygmy and giant regions to the
overall $m_{{}_{-1}}$ moment for ${}^{54}$Ca. We recall that the electric dipole
polarizability, which is proportional to the $m_{{}_{-1}}$ moment of the isovector
dipole response, was identified as a strong isovector indicator that is strongly
correlated to both the neutron-skin thickness of neutron-rich nuclei and the slope 
of the symmetry energy\,\cite{Reinhard:2010wz}. As indicated in Table\,\ref{Table5}
and illustrated in Fig.\,\ref{Fig7},  we too find a correlation between the neutron-skin 
thickness of the neutron-rich calcium isotopes and the total $m_{{}_{-1}}$ moment 
of the isoscalar monopole response. As already alluded in earlier references, see 
for example Refs.\,\cite{Chen:2014mza} and\,\cite{Todd:2003xs}, the softer the 
symmetry energy the earlier the neutron-drip line is reached. In essence, models 
with thin neutron skins predict weak binding energies for the valence neutron 
orbitals in neutron-rich nuclei. In the present case, FSUGarnet having the softest 
symmetry energy of the models employed here (see Table\,\ref{Table2}) predicts
the weakest binding for the valence 2p-1f$^{\,5/2}$ neutron orbitals in the neutron-rich 
calcium isotopes and, as a consequence, the largest amount of low-energy monopole 
strength. In turn, NL3 having the stiffest symmetry energy displays the opposite 
trend: whereas it predicts the thickest neutron skins, its displays the least amount 
of low-energy monopole strength.

\begin{figure}[ht]
 \vspace{-0.1cm}
 \begin{center}
  \includegraphics[width=0.60\linewidth,angle=0]{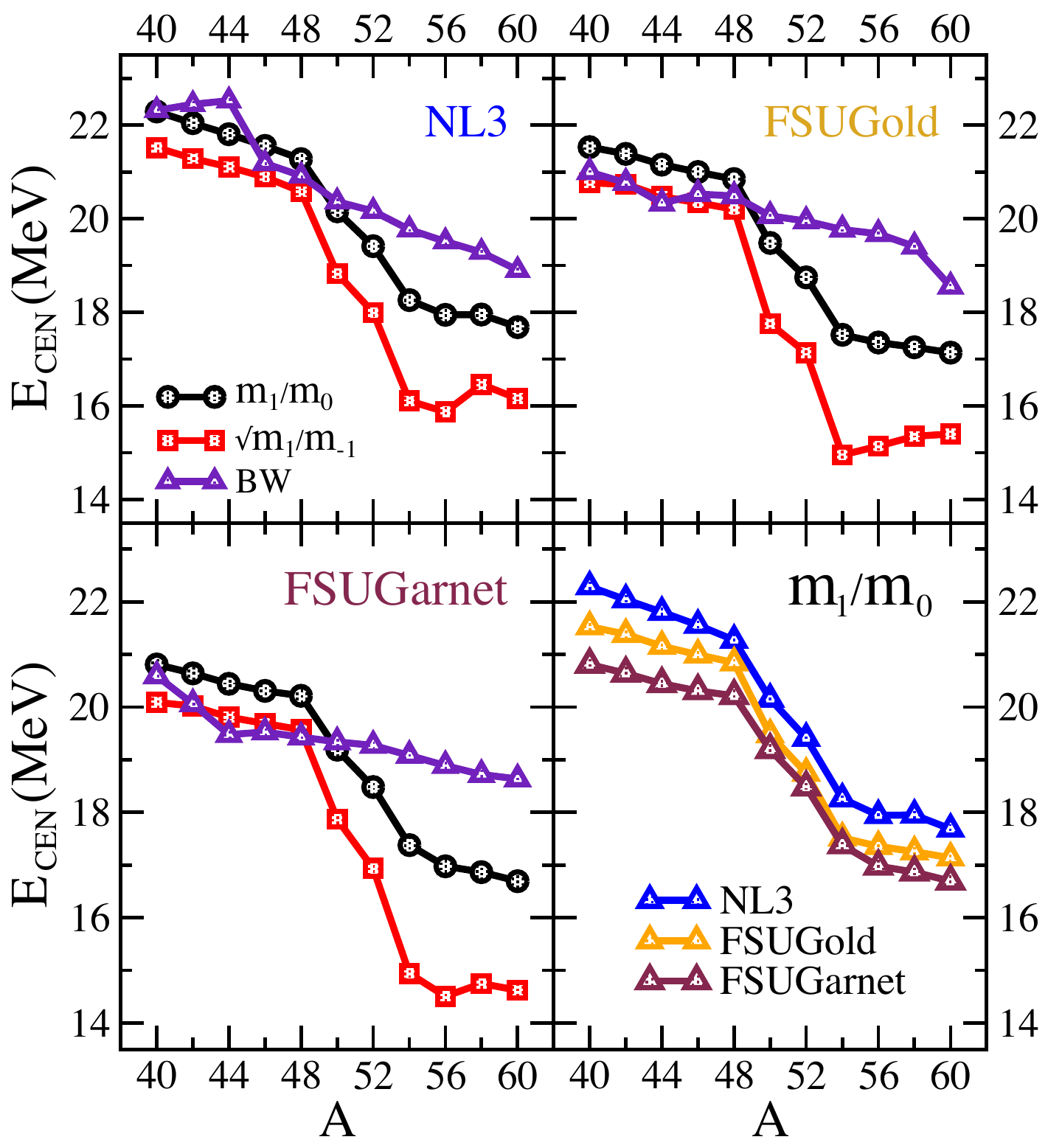}
  \caption{(Color online) Centroid monopole energies predicted by all three 
  relativistic models NL3, FSUGold, and FSUGarnet for all even-even calcium 
  isotopes from ${}^{40}$Ca to ${}^{60}$Ca. The various definitions adopted 
  for the centroid energy have been discussed in the text.}
 \label{Fig8}
 \end{center} 
 \vspace{-0.25cm}
\end{figure}

We conclude this section by displaying in Fig.\,\ref{Fig8} centroid energies for 
all three models considered in the text. Energies have been computed as both 
$m_{{}_{1}}/m_{{}_{0}}$ (black circles) and $\sqrt{m_{{}_{1}}/m_{{}_{-1}}}$
(red squares). Note that $m_{{}_{1}}/m_{{}_{0}}$ values also appear between
parenthesis in both Fig.\,\ref{Fig2} and Fig.\,\ref{Fig5}(a). Besides these two 
choices, we include for reference centroid energies computed by assuming 
a Breit-Wigner (or Lorentzian) fit to only the giant monopole resonance (purple triangles). 
As the strength distribution for all stable calcium isotopes displayed in Fig.\,\ref{Fig2} 
is dominated by a single collective peak, fairly consistent results are obtained 
for all three choices. However, with the appearance of low energy 
monopole strength in ${}^{50}$Ca---and progressively more strength with 
increasing mass number---significant distortions to the simple Lorentzian 
shape emerge. This is reflected in a large dispersion among the three adopted 
definitions. Naturally, the Lorentzian fit to the GMR peak predicts the largest 
energy value, followed by $m_{{}_{1}}/m_{{}_{0}}$, and $\sqrt{m_{{}_{1}}/m_{{}_{-1}}}$ 
predicting the lowest value, because of the large enhancement in $m_{{}_{-1}}$. 
Finally, although the last panel in the figure collects $m_{{}_{1}}/m_{{}_{0}}$ 
predictions from all three models, the notion of a centroid energy for a 
distribution of strength that is no longer dominated by a single collective 
peak loses most of its appeal.

\section{Conclusions}
\label{Conclusions}

The fascinating dynamics of exotic nuclei has lead to a paradigm shift 
in nuclear structure. \emph{``How does subatomic matter organize itself 
and what phenomena emerge?"} and \emph{``What combinations of 
neutrons and protons can form a bound atomic nucleus?"} are only a 
few of the most interesting questions energizing nuclear physics today.
The main goal of this contribution was to explore the nature of the isoscalar 
monopole response in exotic calcium isotopes with extreme combinations 
of neutrons and protons.

At the most basic level, one would like to understand the scaling of the 
giant monopole resonance with increasing mass number, as it encodes 
information on the incompressibility of neutron-rich matter which, in turn, 
is sensitive to both the incompressibility coefficient of symmetric nuclear 
matter and the density dependence of the symmetry energy.  Perhaps 
the biggest challenge in uncovering the sensitivity of the monopole 
resonance to the symmetry energy stems from the fact  that its contribution 
scales as the square of the neutron-proton asymmetry. Thus, one most 
probe the dynamics of exotic nuclei far away from the valley of stability 
where the neutron-proton asymmetry is large. Important first steps along
this direction have been taken along the isotopic chains in tin and 
cadmium. However, given that these measurements have been limited 
so far to stable isotopes, the neutron excess, although appreciable, 
is not yet sufficiently large. Yet, one is confident that in the new era of rare 
isotope facilities experimental studies of this kind will be extended well
beyond stability.
 
An important motivation behind the current work was to confront the 
experimental observation that the distribution of isoscalar monopole 
strength in ${}^{40}$Ca appears softer than in ${}^{48}$Ca, in stark 
contrast with theoretical expectations. To do so, we have used three 
relativistic mean-field models that are known to provide a good 
description of ground-state properties throughout the nuclear 
chart. The disagreement with the experimental trend was confirmed 
for all three models considered here. Indeed, in examining the distribution 
of monopole strength in all stable even-even isotopes from ${}^{40}$Ca 
to ${}^{48}$Ca, we found a gradual, albeit small, softening of the monopole 
response with increasing mass number. This in spite that one of the models 
adopted here (``FSUGarnet") was fitted to GMR energies of several magic 
and semi-magic nuclei. Hence, measurements of the isoscalar monopole 
response of other stable calcium isotopes is strongly encouraged as it may 
play a vital role in resolving this discrepancy.

Yet the central goal of the present work was to study the emergence
evolution, and origin of low-energy monopole strength as a function 
of neutron excess. Somehow surprising, we found no evidence of a 
soft monopole mode in the stable calcium isotopes despite a steady 
increase in the thickness of the neutron skin. However, the situation 
changed dramatically with the occupation of the weakly-bound neutron 
orbitals. Indeed, starting with ${}^{50}$Ca and ending with ${}^{60}$Ca, 
a well developed soft monopole mode of progressively increasing strength 
was clearly identified. The origin of the low-energy monopole strength was 
attributed to neutron excitations from the weakly-bound orbitals into the 
continuum.  As such, the RPA formalism employed here, based on a 
non-spectral approach that treats the continuum on the same footing 
as the bound states, is particularly advantageous. 

Due to the prominent role played by the weakly bound neutron orbitals
in generating low-energy monopole strength, we searched for a 
correlation to the density dependence of the symmetry energy. This is 
motivated by the realization that models with a soft symmetry energy 
reach the neutron-drip line before their stiffer counterparts. In our particular 
case, this is a reflected in the fact that the 2p-1f$^{\,5/2}$ neutron orbitals 
are more weakly bound in FSUGarnet (the softest of the models employed 
here) than in NL3 (the model with the stiffest symmetry energy). As a 
result, FSUGarnet predicted a larger amount of low energy monopole strength 
than NL3.  On the other hand, models with a stiff symmetry energy generate 
thicker neutron skins. Hence, we uncovered the following inverse correlation: 
the thiner the neutron skin  the larger the $m_{-1}$ moment of the monopole
distribution. Given that the isoscalar monopole response of the unstable 
${}^{68}$Ni isotope has already been measured in a pioneer experiment using 
inelastic alpha scattering in inverse kinematics, we are confident that such 
techniques may also be used to explore the monopole response of the exotic
calcium isotopes. Such experiments will provide critical insights into the role 
of the continuum in understanding the physics of weakly-bound systems and 
on the development of novel modes of excitation in exotic nuclei.

\begin{acknowledgments}
The author wishes to express his gratitude to Professor Umesh Garg for 
many stimulating conversations. This material is based upon work supported 
by the U.S. Department of Energy Office of Science, Office of Nuclear Physics 
under Award Number DE-FD05-92ER40750.
\end{acknowledgments}

\bibliography{Ca60IsGMR.bbl}

\end{document}